\documentclass[epj]{svjour}
\usepackage{graphicx,mathrsfs,amsmath,bm}
\usepackage{color}
\newcommand{\gtorder}{\mathrel{\raise.3ex\hbox{$>$}\mkern-14mu
            \lower0.6ex\hbox{$\sim$}}}
\newcommand{\ltorder}{\mathrel{\raise.3ex\hbox{$<$}\mkern-14mu
            \lower0.6ex\hbox{$\sim$}}}

\begin{document}
\title{Neutron star mass limit at $2M_\odot$ supports the existence of a CEP}
%High mass twin stars as indication of the existence of a CEP in the QCD diagram}
\author{
%Mark G. Alford\inst{1}
%\and
D. Alvarez-Castillo\inst{1,}\thanks{\emph{On leave from:} 
Instituto de F\'{i}sica, Universidad Aut\'{o}noma de San Luis Potos\'{i}, M\'{e}xico}
\and
S. Benic\inst{2,}\thanks{\emph{Present address:} University of Tokyo, Tokyo, Japan}
\and 
D. Blaschke\inst{1,3,4} 
\and
Sophia Han\inst{5,6}
\and
S. Typel\inst{7}
}
% etc
\offprints{}          % Insert a name or remove this line
\institute{
%Physics Department, Washington University, St. Louis MO, USA
%\and
Bogoliubov Laboratory of Theoretical Physics, JINR Dubna, Dubna, Russia
\and 
Department of Physics, University of Zagreb, Zagreb, Croatia
\and
National Research Nuclear University (MEPhI), Moscow, Russia
\and
Institute of Theoretical Physics, University of Wroclaw, Wroclaw, Poland
\and
Department of Physics and Astronomy, University of Tennessee, Knoxville, TN 37996, USA 
\and 
Physics Division, Oak Ridge National Laboratory, Oak Ridge, TN 37831, USA
\and
GSI Helmholtzzentrum f\"ur Schwerionenforschung GmbH, Darmstadt, Germany
}
\date{Received: date / Revised version: date}
\abstract{We point out that the very existence of a "horizontal branch" in the mass-radius characteristics for 
neutron stars indicates a strong first-order phase transition and thus supports the existence of a critical endpoint (CEP) of first order phase transitions in the QCD phase diagram. 
This branch would sample a sequence of hybrid stars with quark matter core, leading to the endpoint of stable compact star configurations with the highest possible baryon densities. Since we know of the existence of compact stars with $2M_\odot$, this hypothetical branch has to lie in the vicinity of this mass value, if it exists. We report here a correlation between the maximal radius of the horizontal branch and 
the pressure at the onset of hadron-to-quark matter phase transition which is likely to be a universal quantity of utmost relevance to the upcoming experiments with heavy-ion collisions at NICA and FAIR. 
\PACS{
      {97.60.Jd}{Neutron stars}   \and
      {26.60.Kp}{Equations of state for neutron star matter}   \and
      {12.39.Ki}{Relativistic quark model}
     } % end of PACS codes
} %end of abstract
\maketitle
\section{Introduction}
\label{sec:intro}
The study of the internal composition of neutron stars (NS) is an active field of research which relies on astrophysical observations that allow to refine theoretical models. 
Matter inside neutron star interiors is so dense that nuclear interactions must be taken into account. Fundamental properties like the maximum neutron star mass and 
radius are determined by their internal composition, namely the equation of state (EoS).
Recent observations of massive NS
\cite{Demorest:2010bx,Fonseca:2016tux,Antoniadis:2013pzd} resulted not only in constraints on the stiffness of the EoS, but also in a reduction of the densities reached in the center of a massive NS.
The stiffer the high-density EoS, the higher the maximum mass supported by it and the flatter the density profile of the compact star interior. 
Furthermore, the possibility of deconfined quark matter content inside NS
can be an indication for a critical end point (CEP) in the QCD phase diagram if the deconfinement process occurs via a strong first phase transition. 
In the other hand, in order to prove the existence of a CEP in the QCD phase diagram, it is sufficient to demonstrate that at zero
temperature $T=0$ a first order phase transition exists as a function
of the baryochemical potential $\mu$, since it is established knowledge from ab-initio lattice QCD simulations that at $\mu=0$ 
the transition on the temperature axis is a crossover.
Moreover, the appearance of a gap in the mass-radius relationship for compact stars leads to the phenomenon known 
as third family (aka "mass twins") which will imply
that the $T=0$ equation of state of compact star matter exhibits a
strong first order transition with a latent heat, that for the EoS scheme presented in~\cite{Alford:2013aca} satisfies
$\Delta\varepsilon/\varepsilon_{\rm trans} > 0.6$, where $\varepsilon_{\rm trans}$ is the critical energy density and $\Delta\varepsilon$ the latent heat of the transition.
Since such a strong first order transition under compact
star conditions will remain first order when going to symmetric matter, the observation of a disconnected
branch (also known as third family) of compact stars in the mass-radius ($M-R$) diagram proves the existence of a CEP in QCD.

A very important ingredient in the microscopic description of hadronic matter is the interactions between 
the quarks confined inside baryons as densities increase towards the phase transition. 
These effects are described by an effective excluded volume of baryons that go beyond the simple point-like description and the result is a stiffening of the hadronic EoS. 
Once deconfinement has set in, these quark interactions shall start to increase therefore leading to multiquark interactions 
that will eventually stiffen the quark EoS and will stabilize the third branch in the $M-R$ relation right after the gap that separates the second branch. As we will see in 
the next sections, the high mass twins require an EoS that is stiff both in the hadronic phase,
as well as in the quark phase in order to stabilize the hybrid star solutions, located in the third branch in the $M-R$ diagram.
This result is on contrast to other approaches to quark matter, like CFL solutions in the NJL model~\cite{Klahn:2013kga,Klahn:2006iw}, or holographic descriptions~\cite{Hoyos:2016zke},
where the hadron-quark matter transition leads to an instability against radial oscillations, collapsing the compact star.

A different approach leading to a crossover for hadronic and quark matter is the so called \textit{three window approach}~\cite{Kojo:2015fua}. 
It is characterized by the use of percolated quark matter equations of state at high density where as in the intermediate domain matter is described by an interpolation between the low density hadronic matter
and the high density quark matter. 
The percolation mechanism is the result of the three window domain of quark matter interactions: 
a) few quark exchange at densities below 2$n_0$, 
b) many-quark exchange resulting in a structural change of hadrons at about $2n_0<n<5n_0$ and 
c) baryon overlap namely baryon percolation and Fermi sea development for $n>5n_0$. 
The resulting neutron stars can be as stiff as the observed 2~M$_\odot$, however under this scheme the third branch and the high mass twins will not appear.

It is important to mention that in \cite{Schafer:1998ef} based on the symmetries of the QCD Lagrangian, it has been suggested that cold dense baryonic matter would undergo a continuous transition to quark matter and thus a first order phase transition should not be expected for T=0 matter.
These considerations got supported by \cite{Hatsuda:2006ps}, where it had been found that the Kobayashi-Maskawa-'t Hooft determinant interaction describing the axial anomaly of QCD would correspond by Fierz transformation to an effective repulsive interaction induced by a six-quark vertex coupling to diquark condensates which in effective quark matter models of the NJL type leads to a second critical endpoint at low temperatures, below which the hadron-to-quark matter transition would become a crossover, see also \cite{Abuki:2010jq}.
If, however, the horizontal branch of the neutron star $M-R$ diagram would be observed in nature, these ideas of a quark-hadron continuity must be revised!

We have already underlined the importance of the  high-mass twin stars phenomenon in the framework of the nature of the QCD phase diagram. 
However, there are other aspects that are also covered, namely solutions to the hyperon puzzle~\cite{Baldo:2003vx}, the masquerade problem~\cite{Alford:2004pf}  and the reconfinement case~\cite{Lastowiecki:2011hh,Zdunik:2012dj}.

As for the Astronomy front, upcoming observations from NICER~\cite{nicer} and SKA~\cite{ska} it shall be possible to detect high mass NS twins. 
Bayesian Analysis (BA) studies are useful to assess the possibility of their identification. 
Several results can be found in  \cite{Alvarez-Castillo:2016oln,Ayriyan:2015kit,Alvarez-Castillo:2015via,Alvarez-Castillo:2014nua,arXiv:1408.4449,arXiv:1402.0478}, 
where disjunct M-R constraints have been used for extracting probability measures for hybrid NS EoS.
In the most recent paper~\cite{Alvarez-Castillo:2016oln}, consideration a fictitious radius measurement of both the pure hadronic NS and the corresponding hybrid twin 
has brought interesting results.
By means of BA, the authors performed model discrimination and obtained the most reliable parameters related to the highest probabilities. These results are of great importance to the observers since they provide a guidance for astronomical 
data filtering.

\begin{figure}[!bhtp]
\begin{center}
\includegraphics[width=0.5\textwidth]{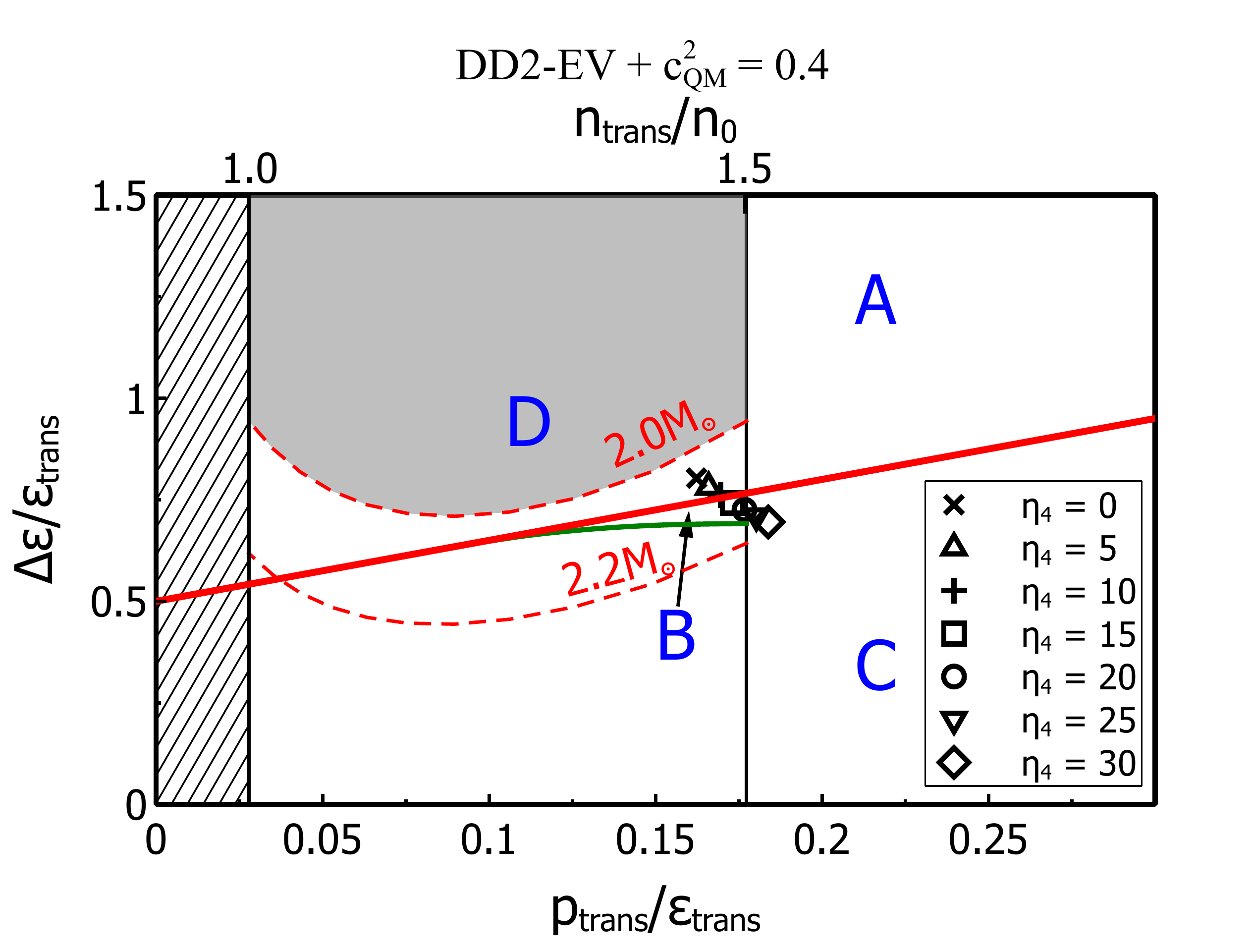}
\end{center} 
\caption{Properties of the hybrid NS EoS. The EoS in the figure corresponds here is composed of the hadronic DD2 that includes excluded volume corrections together with the NJL model with multiquark interactions, see section~\ref{sec:hNJL}.  
The regions are classified according to \cite{Alford:2013aca}, where the $M -R$ configurations have a stable hybrid star branch that can be: 
(D)isconnected from the hadronic branch, ({C})onnected with it, (B)oth or (A)bsent.}
\label{Case_A_n_B}
\end{figure}
This work is organized as follows: In section \ref{sec:hybrid} we review different approaches to the high mass twin EoS, exposing the main features of each model. In section \ref{sec:astro}
we make the connection with Astrophysics by summarizing the most relevant astrophysical cases and observables related to the high mass twins. In section \ref{sec:ns-hic} we discuss the overall
picture and finally we present our results and enumerate the physical inputs for the NICA experiment resulting from the twins phenomenon just before the final conclusions section.

\section{Hybrid compact star EoS models with strong first order phase transition.}
\label{sec:hybrid}
In this section we review several approaches to the dense matter EoS that undergo a strong first-order transition between the hadron and quark regions, describing the high mass twin phenomenon. 
The resulting massive hybrid neutron stars are composed of a hadronic mantle with quark matter cores and 
are located in the third branch of the $M - R$ diagram, where each curve represents sequences of static, non-rotating, compact stars for a given EoS after solution of the TOV equations 
\cite{Tolman:1939jz,Oppenheimer:1939ne}.

The hadronic NS EoS description includes excluded volume corrections that are applied at densities higher than saturation in oder to respect the empirical values obtained from diverse laboratory measurements.

The excluded volume correction is applied at suprasaturation densities and has the effect of stiffening the EoS without modifying any of the experimentally
well constrained properties below and around saturation $n_{\textmd{sat}}=0.16$ fm$^{-3}$, the density in the interior of atomic nuclei.

In~\cite{Alvarez-Castillo:2016oln} the DD2 and DD2F EoS features this correction by considering the available volume fraction $\Phi_N$ for the motion of nucleons as density dependent $n$ as introduced in~\cite{Typel:2016srf}
 \begin{equation}
    \Phi_N=\left\{
                \begin{array}{lll}
                  1~,& \textmd{if} &n \leq n_{\textmd{sat}}\\
                  \exp[-{v\vert v \vert}(n-n_{\textmd{sat}})^{2}/2]~, & \textmd{if} &n > n_{\textmd{sat}}~,
                \end{array}
              \right.
              \label{vex}
  \end{equation}
where $v=16\pi r_N^3/3$ 
is the van-der-Waals excluded volume for a nucleon with a hard-core radius $r_N$. To complete the hybrid NS EoS, a quark matter (QM) EoS is to be included through a phase transition, therefore in the following subsections 
a brief description of the different approaches to the QM EoS is presented.

\subsection{ZHAHP scheme}
This approach~\cite{Zdunik:2012dj,Alford:2013aca} is based on an EoS ansatz for the description of quark matter, characterized by a constant speed of sound $c_{QM}$. 
While keeping the hadronic equation of state
fixed, the strength of the first order phase transition $\Delta \varepsilon$ is varied and the resulting $M - R$ curves are studied.
The EoS takes the following form:
\begin{equation}
\varepsilon(p) = 
\left\{ \begin{array}{ll} 
\varepsilon_{H}(p) \ , & p <  p_{\textrm{trans}} \\
\varepsilon_{\textrm{trans}} +\Delta \varepsilon + c^{-2}_{QM}(p-p_{\textrm{trans}}) \ , &  p > p_{\textrm{trans}} \end{array} 
\right.
\label{eq:eoscross}
\end{equation}
where $\varepsilon_{H}(p)$ is the hadronic EoS, $p_{\rm trans}$ is the pressure at the phase transition
and $\varepsilon_{\textrm{trans}}=\varepsilon_{H}(p_{\textrm{trans}})$ is the energy density at the 
onset of the phase transition. 
Variations of the three parameters $p_{\textrm{trans}}$, $\Delta \varepsilon$ and $c_{QM}$
lead to different sequences of compact star configurations in the $M - R$ diagram. 
According to \cite{Alford:2013aca} the $M -R$ configurations have a stable hybrid star branch that can be: 
(D)isconnected from the hadronic branch, ({C})onnected with it, or (B)oth, 
see Fig.~\ref{Case_A_n_B}.
The line 
\begin{equation}
 \frac{\Delta \varepsilon}{\varepsilon_{\textrm{trans}}} = \frac{1}{2}+ \frac{3}{2} \frac{p_{\textrm{trans}}}{\varepsilon_{\textrm{trans}}}
\end{equation}
is shown in that figure and divides the cases (D) and ({C}). 
Sequences on that line exhibit an endpoint of hadronic configurations from which a "horizontal branch" 
of almost unstable hybrid stars begins.
Such a horizontal branch would be a clearly detectable observational indication for a strong first order phase transition (since $\Delta \varepsilon > \varepsilon_{\textrm{trans}}$) which provides evidence for the existence of a CEP in the QCD phase diagram~\cite{Alvarez-Castillo:2013cxa,Alvarez-Castillo:2015xfa}.

\subsection{Microscopic models of quark matter}

In this section we provide a few examples of microscopic approaches to the quark matter EoS
for which high-mass twin star solutions have been reported.
They all belong to the meanfield level of description for which the grand canonical potential takes the generic form
\begin{eqnarray}
\Omega &=&U+\Omega^{FG}-\Omega_0~,
\label{eq:pot}
\end{eqnarray}
where $U$ is a meanfield contribution and $\Omega_0$ specifies the normalization of the thermodynamical potential of the vacuum.
$\Omega^{FG}=\sum_{f=u,d,s}\Omega^{FG}_f(T,\tilde{\mu}_f;X)$ is the Fermi gas expression which can still depend on order parameters $X$ to be determined in thermodynamic equilibrium from the solutions of the corresponding gap equations $\partial \Omega/\partial X = 0$.
These fix, e.g., the masses and pairing gaps for quarks in dense matter.
The EoS are then obtained the standard way
\begin{equation}
 p=-\Omega, \quad -\frac{ \partial \Omega}{\partial \mu_f}=n_f, \quad \varepsilon=-p+\sum_{f=u,d,s}\mu_f n_f. 
\end{equation}
After implementing the constraints of $\beta-$equilibrium and charge neutrality the EoS will depend only on one chemical potential which can be chosen as that corresponding to the baryon number.
\subsubsection{Bag model with color superconductivity}
\label{sec:cfl}

An interesting hybrid NS EoS that includes color superconducting quark matter in the color flavor locked (CFL) phase has been given in \cite{Alford:2004pf} and was  reported to posses high-mass twin star solutions in~\cite{Agrawal:2009ad,Agrawal:2010er}. 

At high densities, the CFL quark matter phase
expected to be found in the core of hybrid stars.
It is described within the  Nambu-Jona-Lasinio (NJL) model augmented with perturbative QCD corrections and augmented with a background pressure.
The grand canonical thermodynamic potential that takes the generic form of~(\ref{eq:pot}) is composed of the following expressions:
\begin{eqnarray}
 \Omega^{FG}&=&\sum_{f=u,d,s}\frac{3}{\pi^{2}}\int_{0}^{\sqrt{\mu_i^{2}-m_i}}dp p^{2}\left( \sqrt{p^{2}+m_i^{2}}-\mu_i \right),
\end{eqnarray}
where the mass of the light quarks ($u$ and $d$) is neglected and $m_s=150$ MeV. There is also a bag constant for the background field
\begin{equation}
 \Omega_0=-B,
\end{equation}
and the mean field is given by
\begin{equation}
 U=\frac{3}{4\pi^{2}}c\mu^{4}-\frac{3\Delta^{2}\mu^{2}}{\pi^{2}},
\end{equation}
where the term proportional to $\mu^{4}$ corresponds to QCD inspired corrections with $c=0.3$. The second term involving the CFL gap parameter $\Delta$ is the lowest order contribution from the CFL condensate formation. 
In this approach, $\Delta$ is introduced by hand instead of being computed from the gap equations, in contrast with the other models, see for instance,~\cite{Blaschke:2005uj}. 
This EoS is complemented with a Goldstone boson condensate that contributes to the total baryonic pressure. 
Moreover, the quark-hadron phase transition features a mixed phase where charge neutrality
is enforced globally as well as the Gibbs conditions.
The resulting neutron star configurations feature a third branch in the $M-R$ diagram as well as unstable quark configurations that bifurcate from the hadronic sequence.
Hybrid stars have masses in the range $1.0 - 2.1~M_{\odot}$. 
Under this scheme, the 2$M_{\odot}$ NS is realized with the following choice of parameter values 
\begin{equation}
\Delta=50~\textrm{MeV}, \qquad B^{1/4}=218~\textrm{MeV}. 
\nonumber
\end{equation}
The hadronic phase EoS implemented in this model is based on the extended FTRMF model~\cite{Prakash:1995uw} that includes contribution from self-interactions and mixed terms for mesons up to fourth order, but \textit{it does not} include excluded volume corrections.
\subsubsection{NJL model with multi-quark interactions}
\label{sec:hNJL}
Within this approach, the EoS in the high density phase is computed by means of a NJL model with multiquark interactions~\cite{Benic:2014jia,Benic:2014iaa}.
The thermodynamic potential functions are
\begin{eqnarray}
U &=& 2\frac{g_{20}}{\Lambda^2}(\phi_u^2+\phi_d^2) + 12 \frac{g_{40}}{\Lambda^8}(\phi_u^2+\phi_d^2)^2 \nonumber \\
&&-2\frac{\eta_2 g_{20}}{\Lambda^2}(\omega_u^2 + \omega_d^2)-12 \frac{\eta_4 g_{40}}{\Lambda^8}(\omega_u^2+\omega_d^2)^2~,
\end{eqnarray}
where the following shifts apply
\begin{eqnarray}
M_u &= &m+4\frac{g_{20}}{\Lambda^2}\phi_u+ 16\frac{g_{40}}{\Lambda^8}\phi_u^3 + 16\frac{g_{40}}{\Lambda^8}\phi_u \phi_d^2\,\,\,,
\label{eq:cmass}
\\
\tilde{\mu}_u &=& \mu_u-4\frac{g_{02}}{\Lambda^2}\omega_u- 16\frac{g_{04}}{\Lambda^8}\omega_u^3 - 16\frac{g_{04}}{\Lambda^8}\omega_u \omega_d^2\,\,\,,
\label{eq:tildm}
\end{eqnarray}
where the same relations hold for the down quark quantities by interchanging the indices $d$ and $u$. The parameters of the model are computed by means of finding the extremum value of the thermodynamic potential \eqref{eq:pot} with respect to the mean-fields ($X=\phi_u, \, \phi_d,\, \omega_u,\, \omega_d$), i.~e.
The intensity of the vector channels is quantified by the ratios
\begin{eqnarray}
\eta_2 = \frac{g_{02}}{g_{20}} \, , \qquad
\eta_4 = \frac{g_{04}}{g_{40}}~.
\end{eqnarray}
The parameter $\eta_4$ determines the stiffness of the quark matter EoS at high densities, a very important behavior for the stability of the third branch in the $M - R$ 
diagram.
\begin{eqnarray}
\frac{\partial \Omega}{\partial X} = 0\,\,\,,
\end{eqnarray}
and as usual, the pressure is obtained from the relation $p=-\Omega$ with charge neutrality and $\beta$-equilibrium enforced in the standard way. For more details,
see~\cite{Benic:2014jia}.
The hadronic part of the EoS is taken to be the DD2 model with excluded volume correction~\cite{Benic:2014jia} and a Maxwell construction for the hadron-quark
phase transition is implemented. High mass twins are found for $\eta_2=0.08$, $\eta_4=5,6,7,8,9$ and $v=6-8~\textrm{fm}^{3}$({\textrm{p60} - {\textrm{p80}).
The lower row of figure \ref{EoS-MR2} shows high mass NS twins together with the corresponding parameter set. 
\begin{figure*}[!bhtp]
%\begin{center}$
%\begin{array}{cc}
\includegraphics[width=0.6\textwidth]{EoS-reTwins_APR_2} \hspace{-2cm}
\includegraphics[width=0.6\textwidth]{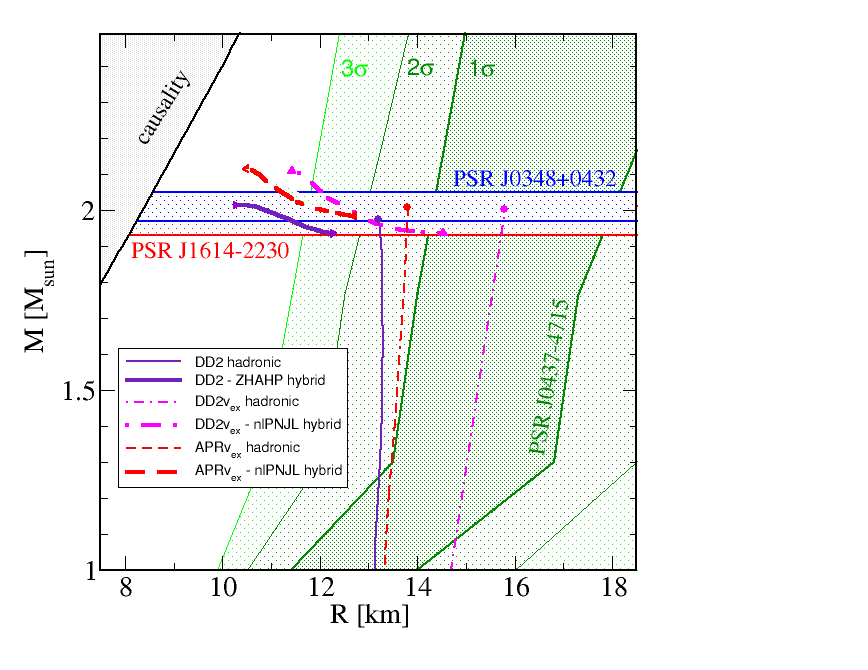}\\
\includegraphics[width=0.6\textwidth]{pvsepsilon_dd2fm_eta4_2Msun} \hspace{-2cm}\includegraphics[width=0.6\textwidth]{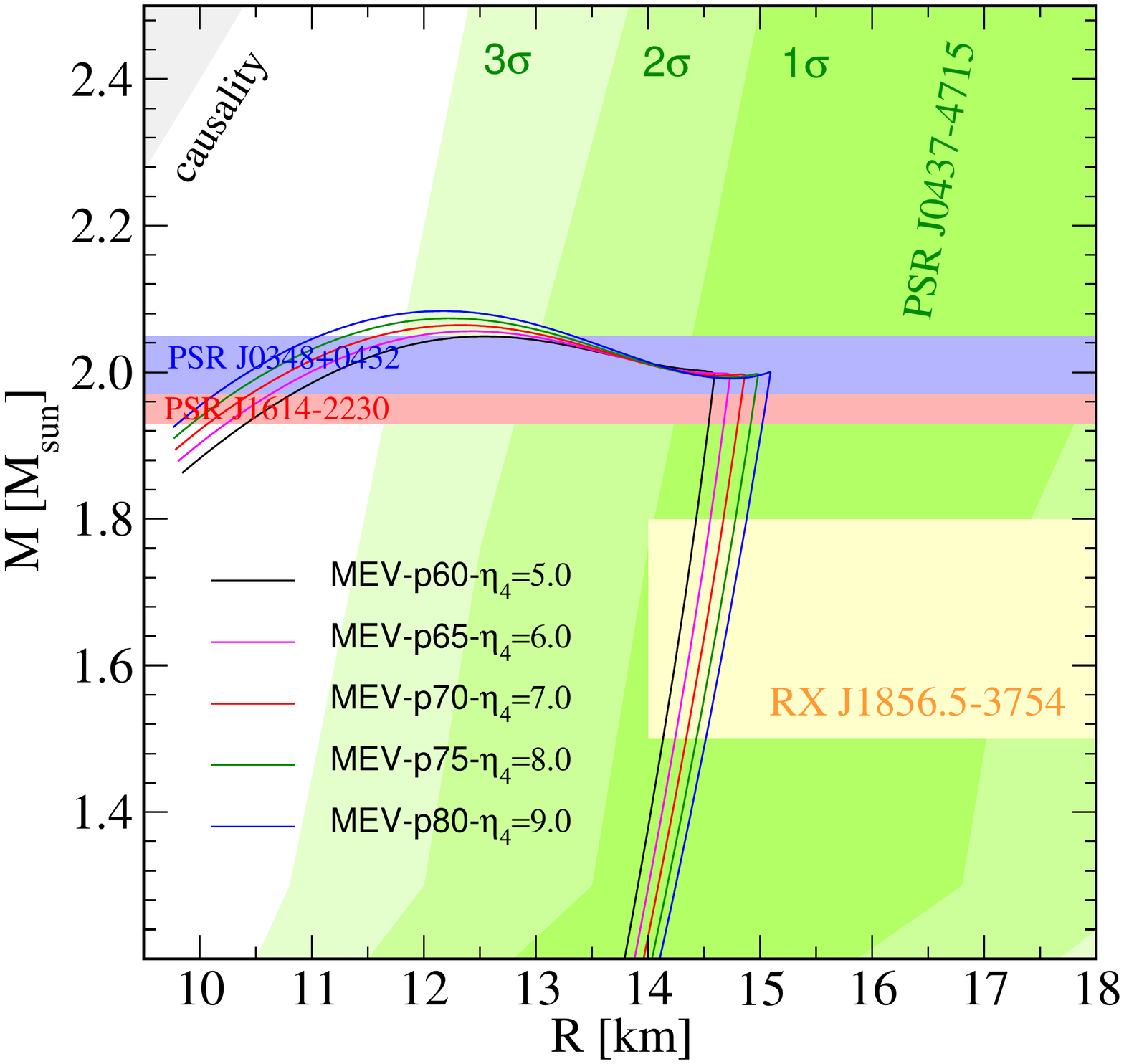}
%\vspace{1cm}
%\end{array}$
%\end{center}
\caption{\label{EoS-MR2}
Left column: High mass NS twins EoS.
Right column:
Corresponding mass-radius diagram for the high mass NS twins EoS. 
The almost vertical branch of purely hadronic stars
is separated from the almost horizontal hybrid star branch by an unstable branch. 
Different approaches to the high mass twins are presented in the upper figures, 
whereas the lower ones display a variation of parameters for the DD2-MEV with hNJL~\cite{Benic:2014jia}.
The recently measured masses for two 2 M$_\odot$ pulsars are shown by red and blue 
hatched regions. Furthermore, the pressure at the phase transition $p_{\textrm{trans}}$ 
on the EoS diagram is proportional to the NS radius in the MR plots, as can be read out from the figures.} 
\end{figure*}
\subsubsection{Non-local quark matter model}
The nonlocal PNJL model~\cite{Contrera:2012wj,Benic:2013eqa} describing the quark phase features a 4-momentum dependent form factor 
that is adjusted to describe the dynamical mass function as well as the wave function renormalization
of the zero temperature propagator based on lattice QCD 
simulations \cite{Parappilly:2005ei,Kamleh:2007ud}. Moreover, the vector meson coupling $\eta_v$ is set to reproduce the
slope of the chemical potential dependence of the pseudocritical temperature
$T_c(\mu)$ as obtained by lattice QCD calculations~\cite{Kaczmarek:2011zz}.
Unlikely the other approaches, for the non-local model the Fermi-Matsubars sums cannot be computed analytically, therefore the mean fermion contribution reads
\begin{eqnarray}
\Omega^{\rm FG} =
- {4 T} \sum_{n,c} \int \frac{d^3\vec p}{(2\pi)^3} \, \,
\mbox{ln} \left[ \frac{ (\tilde{\rho}_{n, \vec{p}}^c)^2 + 
M^2(\tilde{\rho}_{n,\vec{p}}^c)}{Z^2(\tilde{\rho}_{n, \vec{p}}^c)}\right],
\end{eqnarray}
with
\begin{eqnarray}
M(p) & = & Z(p) \left[m + \sigma_1 \ g(p) \right] ~,~~%\nonumber\\
Z(p)  =  \left[ 1 - \sigma_2 \ f(p) \right]^{-1}~. \nonumber
\label{mz}
\end{eqnarray}
Here, instead of a four momentum variable the quantity
\begin{equation}
\Big({\tilde{\rho}_{n,\vec{p}}^c} \Big)^2 =
\Big[ \omega_n - i \tilde{\mu} + \phi_c \Big]^2 + {\vec{p}}\ \! ^2 \ ,
\label{eq:rho}
\end{equation}
occurs which contains the Fermi-Matsubara frequencies $\omega_n=(2 n +1 )\pi T$, the chemical potential $\mu$ and the Polyakov loop variable $\Phi_c$, where
\begin{equation}
\tilde{\mu} = \mu \; - \omega \; g(p) \; Z(p)~.
\label{mutilde}
\end{equation}
For details, see~\cite{Blaschke:2013rma}.
The mean field potential contains scalars ($\sigma_1$,$\sigma_2$) and vector ($\omega$) gaps as well as a coupling to 
the traced Polyakov loop $\Phi_c$ which represents the homogeneous gluon background field $\phi_3$.
\begin{eqnarray}
U=\frac{\sigma_1^2 + \kappa_p^2\ \sigma_2^2}{2\,G_S} - \frac{\omega^2}{2 G_V}\
+ \mathcal{U}(\Phi,T,\mu)~, 
\label{granp}
\end{eqnarray}
with a $\mu$-dependent logarithmic effective potential included:
\begin{eqnarray}
\mathcal{U}(\Phi,T,\mu)&=&(a_0T^4+a_1\mu^4+a_2T^2\mu^2)\Phi^2 \nonumber \\ 
&+& a_3T_0^4\ln{(1-6\Phi^2+8\Phi^3-3\Phi^4)}~,
\label{PL_pot}
\end{eqnarray}
for the parameter values $a_0=-1.85$, $a_1=-1.44\times 10^{-3}$, $a_2=-0.08$,
$a_3=-0.40$. Polyakov loop potential $\mathcal{U}$ contains an explicit dependence on the chemical potential $\mu$ so that an effect like the 
renormalization of the kinetic pressure term by perturbative QCD effects (proportional to the constant $c$ in section~\ref{sec:cfl} is effectively accounted for. 

In addition, the $\mu$ dependence of $\eta_v$ is introduced by interpolating between zero
temperature quark matter pressures $P_< =P_Q(\mu;\eta_<)$ and $P_> =P_Q(\mu;\eta_>)$ 
(under $\beta-$ equilibrium conditions) which are calculated at different, but fixed values 
$\eta_< = \eta_v(\mu \le \mu_c)$ and $\eta_> = \eta_v(\mu > \mu_c)$~\cite{Blaschke:2013ana}.
From the Gibbs conditions, $P_H(\mu_c)=P_<(\mu_c)$, the critical chemical potential $\mu_c$ can be determined.
The interpolation for hybrid NS matter is introduced in a Gaussian form
\begin{equation}
\label{interpol}
 \begin{array}{ll}
P(\mu)=P_H(\mu)\Theta(\mu_c-\mu)+& \\
\left[\left(P_< - P_>\right)
\exp\left[-(\mu-\mu_c)^2/\Gamma^2\right]+P_>\right]\Theta(\mu-\mu_c).&
\end{array}
\end{equation}
where $P_H(\mu)$ stands for the pressure of the hadronic phase in $\beta-$equilibrium. 
As with the other model approaches, the density jump at the first order phase transition is given by
$\Delta n = \partial P_</\partial \mu |_{\mu=\mu_c}
- \partial P_H/\partial \mu |_{\mu=\mu_c}$.
Note that with the choice $\eta_< < \eta_>$ it is possible to achieve a strong first order
phase transition characterized by a large $\Delta n$ allowing for stable hybrid star
configurations at high central energy NS densities.
High mass NS twins examples within this approach are shown in Fig.~\ref{EoS-MR2} (upper row) together with the ZHAHP scheme parametrization of Ref.~\cite{Alvarez-Castillo:2013cxa}. 
Thus, the observation of the massive twins phenomenon would provide evidence for a strong 
first order phase transition in compact star matter. 
\subsubsection{Density functional approach (string-flip model)}
Hybrid compact stars can be described by a excluded volume dependent EoS both in the hadronic and quark matter range.
The string-flip model firstly introduced in~\cite{Horowitz:1985tx,Ropke:1986qs} allows for a
description of quark-nuclear matter based on
the confinement potential for the interactions between nearest neighbor quarks.	
In this section we present the approach developed in~\cite{Alvarez-Castillo:2015rwi} which follows its relativistic form as introduced 
in~\cite{Khvorostukin:2006aw} whose main ingredient is the reduction and vanishing of the effective string tension in dense matter, based on the multiquark interactions prescription. A two-flavor model is introduced with density dependence of the quark masses 
$m_{\textrm{I}}=\Sigma_{\textrm{s}}$, generated by the scalar quark selfenergy $\Sigma_{\textrm{s}}$ (divergent as $n \rightarrow 0$) in order to produce deconfinement.  
Within the density functional approach (string-flip model),
the zero temperature functional for the effective mass takes the form
\begin{equation}
	M_{f} = M_{f_{0}} + D(n_\textrm{s}) n_{\textrm{s}}^{-\frac{1}{3}},
\end{equation}
where the last term is the confinement contribution. The function $D(n_\textrm{s})$ is the effective in-medium string
tension resulting from the product of the vacuum string tension, $D_0$ between quarks with an available volume fraction, $\Phi(n_\textrm{s})$:
\begin{equation}
D(n_\textrm{s})=D_0 \Phi(n_\textrm{s}), \qquad \Phi(n_\textrm{s}) = e^{- v^{2} n_\textrm{s}^{2}},
\end{equation}
where $v$ is the excluded volume parameter. The chemical potential is shifted as follows:
\begin{equation}
 \tilde{\mu}_f=\mu_f-a n_v - b n^{3}_v,
\end{equation}
where $n_v$ is the vector density. The mean field potential in this approach is given by
\begin{equation}
 U=  \frac{1}{3} D(n_\textrm{s})n_{\textrm{s}}^{\frac{2}{3}} -2 v^{2}  D(n_\textrm{s}) n_{\textrm{s}}^{\frac{8}{3}}+a n_{\textrm{v}}^{2}+3n_{\textrm{v}}^{4}.
\end{equation}
The first two terms arise from the string-flip model that includes available volume corrections while the two last ones represent repulsive multiquark interactions.

At lower densities the confinement interaction is expected to dominate, while the perturbative one gluon exchange interaction turns out to be more important at higher densities. 
A very important ingredient of this string-flip model is the effective in-medium string tension $D(n_B)$, which is the result of the product of the vacuum string 
tension $\sigma\sim D_0$ between quarks with the available volume fraction $\Phi(n_\textrm{B})$.
In this way, the excluded volume mechanism is active in both hadron and quark dense matter phases.
For the hadronic phase the DD2 model with excluded volume corrections is taken and the Maxwell construction is applied. High mass twin stars sequences are 
obtained in~\cite{Alvarez-Castillo:2015rwi} for the parameters given there.
\section{Mass-radius relations for high mass twins}
\label{sec:astro}

$M -R $ measurements are important not only because it can be used as an indicator of hybrid compact stars, but of the existence of a CEP in the QCD diagram. 
In particular, the confirmation of a third  branch disconnected from the hadronic neutron star branch suggested by several high mass twin models would support the existence of a first order phase transition, expected to be deconfinement. 
Due to limited accuracy of mass and radius measurements, such a third branch would reveal itself as a quasi horizontal part of the M-R relation at high mass.
There exists also the possibility of the appearance of pasta phases at the interface between hadronic and quark matter in NS star interiors. 
A clear example is found in~\cite{Yasutake:2014oxa}, where the authors consider different structural shapes at the boundary. 
Therefore, in~\cite{Alvarez-Castillo:2014dva} a study has been carried out with the purpose of testing the robustness of the high mass twins against pasta content by 
using an interpolation whose main parameter characterizes the pasta strength and it was found that for moderate values the twin
phenomena remains. Thus, this study provides support for the robust existence of the high mass NS twins. 

From the phenomenology of the massive hybrid stars and the third branch we can point out the following:

\begin{list}{\labelitemi}{\leftmargin=0.5cm}
\item[(1)]  The hadronic phase should be stiff enough to reach the 2$M_{\odot}$ measured NS. Excluded volume corrections can effectively account for it by stiffening the EoS.
\item[(2)] The quark matter EoS can be soft around the quark-hadron phase interface region, especially when resulting in a strong first order phase transition, but should be stiffen up as density increases in order
to maintain the stability of the third branch and result in the case D of figure~\ref{Case_A_n_B}.
\end{list}

%Due to nature of the twins scenario, the transition between twin stars bears an energy reservoir \cite{Alvarez-Castillo:2015dqa,Alvarez-Castillo:2016dyz}
%that qualifies it as a possible engine for most energetic explosive astrophysical phenomena like gamma-ray bursts and fast radio bursts or play a role in contributing to the complex mechanism of core collapse supernova explosions. 
%Furthermore, there are other astrophysical observables for the twins that include NS cooling histories, NS spin-up and spin-down, gravitational wave and neutrino emissions, to name a few ones.

\section{Interplay between neutron star physics and heavy ion collisions}
\label{sec:ns-hic}

High mass NS twins scenario has been successfully described by different microscopic approaches to the EoS. We look into common features in this section. The EoS is characterized by a strong first order phase transition
accompanied by a latent heat. The hadronic EoS is generally stiff while the quark matter description usually softens right after the transition and stiffens up at higher densities, providing the necessary stability for the
compact stars that populate the third branch in the MR diagram.

From statistical models of particle production in heavy ion collisions, a {universal transition pressure} $p_{\rm trans}=82 \pm 8~$MeV/fm$^{3}$ for the quark-gluon plasma (QGP)
breakup and chemical freezeout is given in~\cite{Petran:2013qla,Rafelski:2014fqa}. 
This study 
covers the entire range of today's accessible reaction energies, with $\sqrt{s_{NN}}$ ranging from $62.4$ GeV to $2760$ GeV . 
By looking at Fig.~\ref{pvs-R-correlation} we can associate a NS radius of $13.8$ km for a high mass twin model with $p_{\rm trans}=80$ MeV/fm$^{3}$.
Fig.~\ref{pvs-R-correlation} shows a reciprocate relation between the NS radius and the pressure $p_{\rm trans}$ at the onset of quark matter in NS cores using the data from Fig.~\ref{EoS-MR2}. 

The fit formula for the relation between radius at $2M_\odot$ and transition pressure $p_{\rm trans}$ is
\begin{eqnarray}
\frac{R}{10~{\rm km}} &=& 1.495 - 0.0021\left(\frac{p_{\rm trans}}{{\rm MeV/fm}^3}-22\right)\nonumber\\
&&+\left(\frac{p_{\rm trans}}{{\rm MeV/fm}^3}-31\right)^{-1}.
\end{eqnarray}
An important quantity that related the EoS of symmetric matter with the NS EoS is the symmetry energy $E_{s}$, best determined by laboratory experiments around saturation density values than at high densities regions where at the present remains quite uncertain. 
The relevance of $E_{s}$ for dense matter relies on the fact that it has a huge impact on the NS radius~\cite{AlvarezCastillo:2012rf}, allowing for a link between astrophysical observations and laboratory experiments.
It has been recently conjectured in~\cite{Blaschke:2016lyx}, that a \textit{universal symmetry energy contribution to the NS EoS} exists, leading to the possibility of directly extracting the symmetric matter EoS from NS observations and vice versa.

\begin{figure}[!bhtp]
\includegraphics[width=0.52\textwidth]{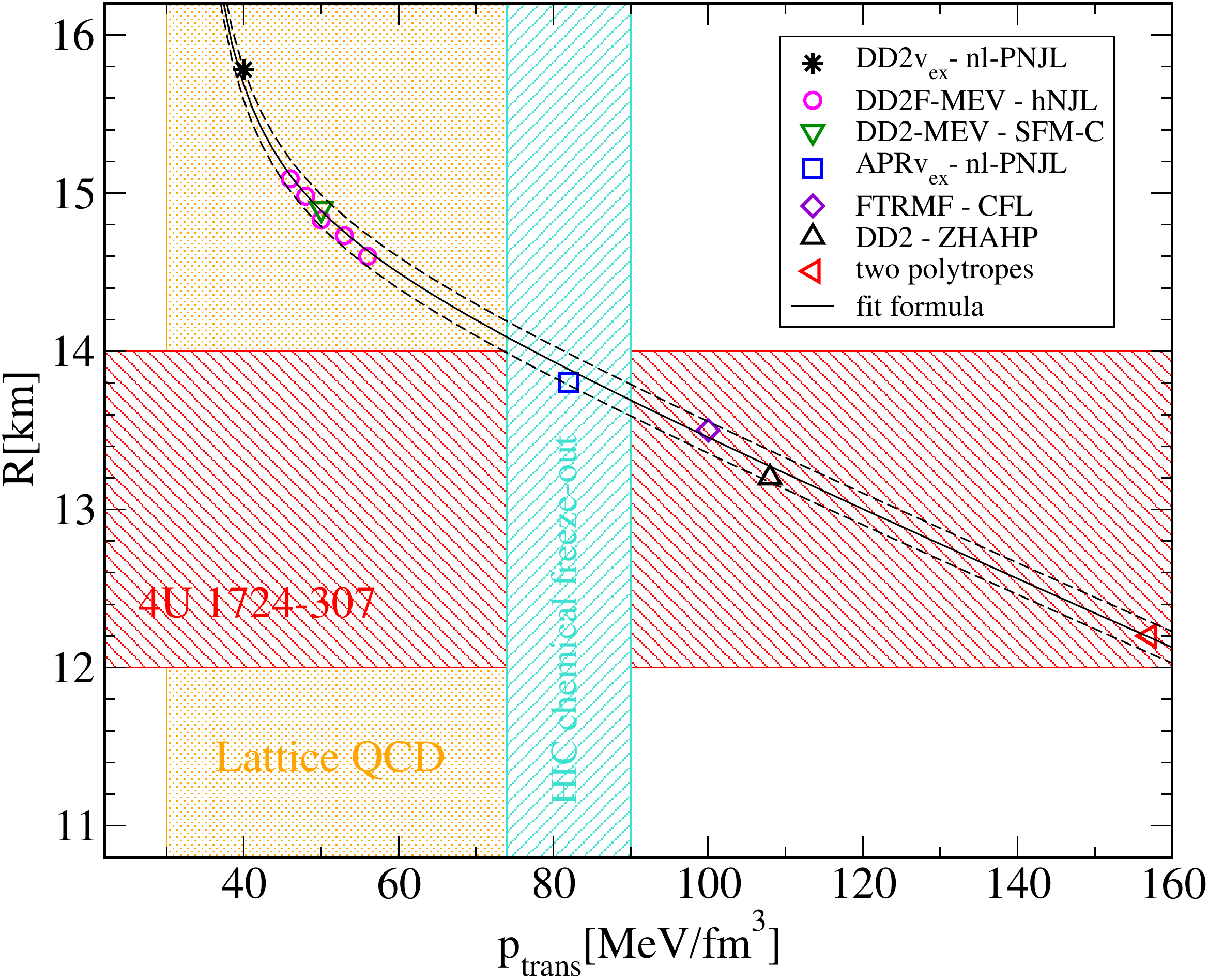}
\caption{\label{pvs-R-correlation}
NS Radius dependence on the $p_{\rm trans}$ for the models presented in figure~\ref{EoS-MR2}.  
The square, the star and triangle-up symbols correspond to the models presented in~\cite{Blaschke:2013ana}, where as the circles represent parameter values of the mass twin configurations in the framework of~\cite{Benic:2014jia} extracted from Ref.~\cite{Alvarez-Castillo:2016oln}.
Additionally, the diamond symbol is the result from the superconducting quark matter model~\cite{Agrawal:2009ad}, the triangle-down symbol stands for the string-flip model approach~\cite{Alvarez-Castillo:2015rwi} and the triangle-left corresponds to the two polytropes EoS~\cite{Bejger}.
The shaded region represent results from Lattice QCD calculations for the crossover transition region~\cite{Bazavov:2014pvz}, from a statistical model analysis of chemical freeze-out in heavy ion collisions that indicate a universal transition pressure~\cite{Petran:2013qla,Rafelski:2014fqa} and from a radius analysis for the X-ray burst spectra for the low-mass X-ray binary 4U 1724-307 \cite{Suleimanov:2015fwa}. 
Remarkably, the model with a radius $R_{\rm trans}=13.8$ km at onset of deconfinement is consistent 
with the universal transition pressure conjecture as well as with the recent radius measurement for 
the LMXB 4U 1724-307.} 
\end{figure}
\section{Conclusions}

From the investigation of the interrelations between the phenomenology of mass-radius relations for compact stars and the underlying equation of state for cold dense matter reviewed in this contribution, 
we can draw a few conclusions applying for the future HIC programmes at NICA, FAIR, low-energy RHIC and the possible heavy-ion collision program at J-PARC:

\begin{list}{\labelitemi}{\leftmargin=0.5cm}
\item[(1)]  If the existence of a disconnected "third family" branch of compact stars in the M-R diagram 
(due to finite errors in the radius and mass measurements hardly distinguishable from a "horizontal branch") would be confirmed by observation, this would allow the conclusion that there must be a strong first order phase transition in the neutron star interior, most likely due to deconfinement.

\item[(2)] A first order phase transition at T=0 in the QCD phase diagram would imply the existence of at least one critical endpoint: worth finding it in HIC experiments!
If even in beta-equilibrium there is a first-order transition, it shall be there also in symmetric matter where there is less stress.

\item[(3)]  If a maximal radius of high-mass pulsars could be established this would provide a lower limit for the pressure at the onset of the phase transition in neutron star matter. This would provide an important constraint for the EoS models used in simulating and interpreting the heavy-ion collisions.

\end{list}

Already at present, without having observed yet the disconnected branch of high-mass twins in the M-R diagram, we may formulate an interesting conjecture, on the basis of our Fig.~\ref{pvs-R-correlation}. 
Suppose the maximum radius ($\sim14$ km) of the range recently extracted from X-ray bursts of 
the compact star 4U 1724-307 by Suleimanov et al.~\cite{Suleimanov:2015fwa} is such a maximum radius of hadronic compact stars, then the lower limit for the transition pressure would be at 
$\sim 80$ MeV/fm$^3$.
This is in striking agreement with the recently reported "universal" hadronization pressure from analyses of chemical freeze-out in HIC by Rafelski and Petran~\cite{Petran:2013qla}.  
So we might conjecture that there is a universal limiting pressure of hadronic matter in Nature, regardless of the direction in which we might explore the QCD phase diagram: temperature, baryon density or isospin density.

\begin{table}[!th]
\center
\caption{The onset pressure $p_{\textrm{trans}}$ for symmetric matter (s)  and in beta equilibrated NS matter (b) for the DD2-MEV vs. hNJL hybrid EoS, and the relative difference $D=2(p_{\textrm{trans}}^{(s)}-p_{\textrm{trans}}^{(b)})/(p_{\textrm{trans}}^{(s)}+p_{\textrm{trans}}^{(b)})$ .}
\label{TablePressures}
\begin{tabular}{ccccc}
\hline \hline 
$\eta_4$&$v$& $p_{\textrm{trans}}^{(s)}$& $p_{\textrm{trans}}^{(b)}$&   D \\
&[fm$^3$] & [MeV/fm$^{3}$]& [MeV/fm$^{3}$]& $\%$  \\
\hline
5     &   6.0    &    65.54       &   56.01			 &   15.68\\
6     &   6.5  &      61.49      &   52.82			&   15.17\\
7     &   7.0  &       58.13       &   50.38			&  14.28\\
8     &   7.5  &      55.13         &   47.89		  	&  14.04\\
9     &   8.0  &        52.63        &   45.98		         &  13.49\\
\hline \hline
\end{tabular}
\end{table}

Table~\ref{TablePressures} corroborates our conjecture for the DD2-MEV vs. hNJL hybrid EoS. 
The relative difference in the transition pressures $D=2(p_{\textrm{trans}}^{(s)}-p_{\textrm{trans}}^{(b)})/(p_{\textrm{trans}}^{(s)}+p_{\textrm{trans}}^{(b)})$ is at the level of $15\%$ only.
Using this correlation together with the largest measured radius for the compact object 4U 1724-307 we find a $p_{\textrm{trans}}$ of about 
$80~\textrm{MeV}/\textrm{fm}^{3}$ that perfectly agrees with the transition pressure extracted from experimental data using a statistical model at low temperature 
and high density in~\cite{Petran:2013qla}.

\subsection*{Acknowledgements}
%\iffalse
We acknowledge the partial support 
%by the Helmholtz Association (HGF) through the Nuclear Astrophysics Virtual Institute (VH-VI-417) and 
by the COST Action MP 1304 "NewCompStar" for our international networking activities in preparing this article. This work received support from the Polish NCN under grant No. UMO-2014/13/ B/ST9/02621.
%D.E.A-C. and H.G. are grateful for support from the programme for exchange between JINR Dubna and 
%Polish Institutes (Bogoliubov-Infeld programme). \\
D.E.A-\-C. and S.T. received support  form the Heisenberg-Landau programme for scientist exchange between JINR Dubna and German Institutes.
S. B. acknowledges partial support by the Croatian Science Foundation under Project No. 8799. 
D.B. was supported in part by the Hessian LOEWE initiative through HIC for FAIR.
%\fi

%\bibliography{epja}

\begin{thebibliography}{99}


%\cite{Demorest:2010bx}
\bibitem{Demorest:2010bx}
  P.~Demorest, T.~Pennucci, S.~Ransom, M.~Roberts and J.~Hessels,
  %{\it Shapiro Delay Measurement of A Two Solar Mass Neutron Star.}
  Nature {\bf 467}, 1081 (2010).
%  doi:10.1038/nature09466
%  [arXiv:1010.5788 [astro-ph.HE]].
  %%CITATION = doi:10.1038/nature09466;%%

%\cite{Fonseca:2016tux}
\bibitem{Fonseca:2016tux} 
  E.~Fonseca {\it et al.},
  %``The NANOGrav Nine-year Data Set: Mass and Geometric Measurements of Binary Millisecond Pulsars,''
  arXiv:1603.00545 [astro-ph.HE].
  %%CITATION = ARXIV:1603.00545;%%

\bibitem{Antoniadis:2013pzd}
J.~Antoniadis {\it et al.},
%,  P.~C.~C.~Freire, N.~Wex, T.~M.~Tauris, R.~S.~Lynch,
%M.~H.~van~Kerkwijk, M.~Kramer, C.~Bassa, V.~S.~Dhillon, T.~Driebe,
%J.~W.~T.~Hessels, V.~M.~Kaspi, V.~I.~Kondratiev, N.~Langer,
%T.~R.~Marsh, M.~A.~McLaughlin, T.~T.~Pennucci, S.~M.~Ransom,
%I.~H.~Stairs, J.~van~Leeuwen, J.~P.~W.~Verbiest, D.~G.~Whelan,
%{\it A Massive Pulsar in a Compact Relativistic Binary},
  Science {\bf 340}, 6131 (2013).
%  doi:10.1126/science.1233232
%  [arXiv:1304.6875 [astro-ph.HE]].
  %%CITATION = doi:10.1126/science.1233232;%%

  
%\cite{Klahn:2006iw}
\bibitem{Klahn:2006iw} 
  T.~Kl\"ahn, D.~Blaschke, F.~Sandin, C.~Fuchs, A.~Faessler, H.~Grigorian, G.~R\"opke and J.~Tr\"umper,
  %``Modern compact star observations and the quark matter equation of state,''
  Phys.\ Lett.\ B {\bf 654}, 170 (2007)
  %doi:10.1016/j.physletb.2007.08.048
  %[nucl-th/0609067].
  %%CITATION = doi:10.1016/j.physletb.2007.08.048;%%
  %121 citations counted in INSPIRE as of 16 Jul 2016
  
  %\cite{Klahn:2013kga}
\bibitem{Klahn:2013kga} 
  T.~Kl\"ahn, R. Lastowiecki and D.~B.~Blaschke,
  %``Implications of the measurement of pulsars with two solar masses for quark matter in compact stars and heavy-ion collisions: A Nambu–Jona-Lasinio model case study,''
  Phys.\ Rev.\ D {\bf 88}, 085001 (2013)
 % doi:10.1103/PhysRevD.88.085001
 % [arXiv:1307.6996].
  %%CITATION = doi:10.1103/PhysRevD.88.085001;%%
  %44 citations counted in INSPIRE as of 16 Jul 2016

%\cite{Hoyos:2016zke}
\bibitem{Hoyos:2016zke} 
  C.~Hoyos, D.~Rodriguez Fernandez, N.~Jokela and A.~Vuorinen,
  %``Holographic quark matter and neutron stars,''
  Phys.\ Rev.\ Lett.\  {\bf 117}, no. 3, 032501 (2016)
  %doi:10.1103/PhysRevLett.117.032501
  %[arXiv:1603.02943 [hep-ph]].
  %%CITATION = doi:10.1103/PhysRevLett.117.032501;%%
  %3 citations counted in INSPIRE as of 16 Jul 2016
  
    %\cite{Kojo:2015fua}
\bibitem{Kojo:2015fua} 
  T.~Kojo,
  %``Phenomenological neutron star equations of state: 3-window modeling of QCD matter,''
  Eur.\ Phys.\ J.\ A {\bf 52}, no. 3, 51 (2016).
  %doi:10.1140/epja/i2016-16051-0
  %[arXiv:1508.04408 [hep-ph]].
  %%CITATION = doi:10.1140/epja/i2016-16051-0;%%
  %3 citations counted in INSPIRE as of 23 Jun 2016
  
   %\cite{Schafer:1998ef}
\bibitem{Schafer:1998ef} 
  T.~Sch\"afer and F.~Wilczek,
  %``Continuity of quark and hadron matter,''
  Phys.\ Rev.\ Lett.\  {\bf 82}, 3956 (1999).
%  doi:10.1103/PhysRevLett.82.3956
%  [hep-ph/9811473].
  %%CITATION = doi:10.1103/PhysRevLett.82.3956;%%
 
 %\cite{Hatsuda:2006ps}
\bibitem{Hatsuda:2006ps} 
  T.~Hatsuda, M.~Tachibana, N.~Yamamoto and G.~Baym,
  %``New critical point induced by the axial anomaly in dense QCD,''
  Phys.\ Rev.\ Lett.\  {\bf 97}, 122001 (2006).
%  doi:10.1103/PhysRevLett.97.122001
%  [hep-ph/0605018].
  %%CITATION = doi:10.1103/PhysRevLett.97.122001;%%
  
    
  %\cite{Abuki:2010jq}
\bibitem{Abuki:2010jq} 
  H.~Abuki, G.~Baym, T.~Hatsuda and N.~Yamamoto,
  %``The NJL model of dense three-flavor matter with axial anomaly: the low temperature critical point and BEC-BCS diquark crossover,''
  Phys.\ Rev.\ D {\bf 81}, 125010 (2010).
%  doi:10.1103/PhysRevD.81.125010
%  [arXiv:1003.0408 [hep-ph]].
  %%CITATION = doi:10.1103/PhysRevD.81.125010;%%

  
   %\cite{Baldo:2003vx}
\bibitem{Baldo:2003vx} 
  M.~Baldo, G.~F.~Burgio and H.-J.~Schulze,
 %{\it Neutron star structure with hyperons and quarks},
  in: \emph{Superdense QCD Matter and Compact Stars},
 %edited by D.   Blaschke and D. Sedrakian, 
 Springer, Heidelberg, 2006, p.113.
 %[{\tt arXiv:0312446 [astro-ph]}].
  %%CITATION = ASTRO-PH/0312446;%%
 
 %\cite{Alford:2004pf}
\bibitem{Alford:2004pf}
  M.~Alford, M.~Braby, M.~W.~Paris and S.~Reddy,
  %{\it Hybrid stars that masquerade as neutron stars},
  Astrophys.\ J.\  629, 969 (2005).
% [{\tt arXiv:0411016 [nucl-th]}].
  %%CITATION = NUCL-TH/0411016;%%
  
  %\cite{Lastowiecki:2011hh}
\bibitem{Lastowiecki:2011hh} 
  R.~Lastowiecki, D.~Blaschke, H.~Grigorian and S.~Typel,
 % {\it Strangeness in the cores of neutron stars},
  Acta Phys.\ Polon.\ Supp.\  {\bf 5}, 535 (2012).
%  [{\tt arXiv:1112.6430 [nucl-th]}].
  %%CITATION = ARXIV:1112.6430;%%
  
  %\cite{Zdunik:2012dj}
\bibitem{Zdunik:2012dj}
J.L.~Zdunik and P.~Haensel.
%{\it Maximum mass of neutron stars and strange neutron-star cores},
  Astron.\ Astrophys.\  {\bf 551}, A61 (2013).
%  doi:10.1051/0004-6361/201220697
%  [arXiv:1211.1231 [astro-ph.SR]].
  %%CITATION = doi:10.1051/0004-6361/201220697;%%
  
   \bibitem{nicer}
{\tt https://heasarc.gsfc.nasa.gov/docs/nicer}

\bibitem{ska}
{\tt http://www.ska.ac.za}

 %%%%%%%%%%%%%%%%%%%%%%%% %Bayesian Analysis 
 
 %\cite{Alvarez-Castillo:2016oln}
\bibitem{Alvarez-Castillo:2016oln} 
  D.~Alvarez-Castillo, A.~Ayriyan, S.~Benic, D.~Blaschke, H.~Grigorian and S.~Typel,
  %``New class of hybrid EoS and Bayesian M-R data analysis,''
  Eur.\ Phys.\ J.\ A {\bf 52}, no. 3, 69 (2016).
  %doi:10.1140/epja/i2016-16069-2
  %[arXiv:1603.03457 [nucl-th]].
  %%CITATION = doi:10.1140/epja/i2016-16069-2;%%
  %1 citations counted in INSPIRE as of 01 Jun 2016
 
 %\cite{Ayriyan:2015kit}
\bibitem{Ayriyan:2015kit} 
  A.~Ayriyan, D.~E.~Alvarez-Castillo, D.~Blaschke and H.~Grigorian,
  %``Mass-radius constraints for the neutron star EoS - Bayesian analysis,''
  J.\ Phys.\ Conf.\ Ser.\  {\bf 668}, no. 1, 012038 (2016).
  %doi:10.1088/1742-6596/668/1/012038
  %[arXiv:1511.05880 [nucl-th]].
  %%CITATION = doi:10.1088/1742-6596/668/1/012038;%%
  %3 citations counted in INSPIRE as of 01 Jun 2016
 
 %\cite{Alvarez-Castillo:2015via}
\bibitem{Alvarez-Castillo:2015via} 
  D.~E.~Alvarez-Castillo, A.~Ayriyan, D.~Blaschke and H.~Grigorian,
  %``Bayesian analysis for two-parameter hybrid EoS with high-mass compact star twins,''
  eConf C140926 (2015).
  %[arXiv:1506.07755 [astro-ph.HE]].
  %%CITATION = ARXIV:1506.07755;%%
  %3 citations counted in INSPIRE as of 01 Jun 2016
 
 %\cite{Alvarez-Castillo:2014nua}
\bibitem{Alvarez-Castillo:2014nua} 
  A.~Ayriyan, D.~E.~Alvarez-Castillo, D.~Blaschke, H.~Grigorian and M.~Sokolowski,
  %``New Bayesian analysis of hybrid EoS constraints with mass-radius data for compact stars,''
  Phys.\ Part.\ Nucl.\  {\bf 46}, no. 5, 854 (2015).
  %doi:10.1134/S1063779615050044
  %[arXiv:1412.8226 [astro-ph.HE]].
  %%CITATION = doi:10.1134/S1063779615050044;%%
  %7 citations counted in INSPIRE as of 01 Jun 201
  
  %\cite{arXiv:1408.4449}
\bibitem{arXiv:1408.4449} 
D.~Alvarez-Castillo, A.~Ayriyan, D.~Blaschke and H.~Grigorian, 
%Bayesian Analysis of Hybrid EoS based on Astrophysical Observational Data.%
%LIT Scientific Report 2011-2013, JINR Publishing Department, Dubna (2014), 
%pp. 123-126, %Editors: Gh. Adam, V.V. Korenkov, D.V. Podgainy, T.A. Strizh, P.V. Zrelov. %ISBN 978-5-9530-0381-0 [arXiv:1408.4449 [astro-ph.HE]].
arXiv:1408.4449 %[astro-ph.HE]].

%\cite{arXiv:1402.0478}
\bibitem{arXiv:1402.0478} 
D.~B.~Blaschke, H.~A.~Grigorian, D.~E.~Alvarez-Castillo and A.~S.~Ayriyan. 
%Mass and radius constraints for compact stars and the QCD phase diagram.
J.\ Phys.\ Conf.\ Ser.\  {\bf 496}, 012002 (2014).
%[arXiv:1402.0478 [astro-ph.HE]].
  
  %\cite{Tolman:1939jz}
\bibitem{Tolman:1939jz} 
  R.~C.~Tolman,
  %``Static solutions of Einstein's field equations for spheres of fluid,''
  Phys.\ Rev.\  {\bf 55}, 364 (1939).
 % doi:10.1103/PhysRev.55.364
  %%CITATION = doi:10.1103/PhysRev.55.364;%%

%\cite{Oppenheimer:1939ne}
\bibitem{Oppenheimer:1939ne} 
  J.~R.~Oppenheimer and G.~M.~Volkoff,
  %``On Massive neutron cores,''
  Phys.\ Rev.\  {\bf 55}, 374 (1939).
%  doi:10.1103/PhysRev.55.374
  %%CITATION = doi:10.1103/PhysRev.55.374;%%

  
 %%%%%%%%%%%%%%%%%%%%%%%%%%%%%%%%%%%%%% 
  
      %\cite{Typel:2016srf}
\bibitem{Typel:2016srf} 
  S.~Typel,
  %{\it Variations on the excluded-volume mechanism},
  Eur.\ Phys.\ J.\ A {\bf 52}, 16 (2016).
%  doi:10.1140/epja/i2016-16016-3
  %%CITATION = doi:10.1140/epja/i2016-16016-3;%%
  
  
 %\cite{Alford:2013aca}
\bibitem{Alford:2013aca}
  M.~G.~Alford, S.~Han and M.~Prakash,
 % {\it Generic conditions for stable hybrid stars},
  Phys.\ Rev.\ D {\bf 88}, 083013 (2013).
%  doi:10.1103/PhysRevD.88.083013
%  [arXiv:1302.4732 [astro-ph.SR]].
  %%CITATION = doi:10.1103/PhysRevD.88.083013;%%
  

 %\cite{Alvarez-Castillo:2013cxa}
\bibitem{Alvarez-Castillo:2013cxa} 
  D.~E.~Alvarez-Castillo and D.~Blaschke,
  %``Proving the CEP with compact stars?,''
  arXiv:1304.7758 %[astro-ph.HE].
  %%CITATION = ARXIV:1304.7758;%%
  %11 citations counted in INSPIRE as of 23 Jun 2016
  
%\cite{Alvarez-Castillo:2015xfa}
\bibitem{Alvarez-Castillo:2015xfa}
  D.~E.~Alvarez-Castillo and D.~Blaschke,
 % {\it Supporting the existence of the QCD critical point by compact star observations},
   PoS CPOD  2014 (2015) 045.
  %[{\tt arXiv:1503.05576 [astro-ph.HE]}].
  %%CITATION = ARXIV:1503.05576;%%
  
  %%%%%%%%%%%%%%%%%

    %%%%%%%%%%%%%%%%%%%%%%Bag models with color superconductivity
  
  %\cite{Agrawal:2009ad}
\bibitem{Agrawal:2009ad} 
  B.~K.~Agrawal and S.~K.~Dhiman,
  %``Stable configurations of hybrid stars with colour-flavour-locked core,''
  Phys.\ Rev.\ D {\bf 79}, 103006 (2009).
  %doi:10.1103/PhysRevD.79.103006
  %[arXiv:0904.2946 [astro-ph.HE]].
  %%CITATION = doi:10.1103/PhysRevD.79.103006;%%
  %6 citations counted in INSPIRE as of 24 Jun 2016
  
  %\cite{Agrawal:2010er}
\bibitem{Agrawal:2010er} 
  B.~K.~Agrawal,
  %``Equations of state and stability of color-superconducting quark matter cores in hybrid stars,''
  Phys.\ Rev.\ D {\bf 81}, 023009 (2010).
  %doi:10.1103/PhysRevD.81.023009
  %[arXiv:1001.1584 [astro-ph.HE]].
  %%CITATION = doi:10.1103/PhysRevD.81.023009;%%
  %14 citations counted in INSPIRE as of 24 Jun 2016
  
    %\cite{Blaschke:2005uj}
\bibitem{Blaschke:2005uj} 
  D.~Blaschke, S.~Fredriksson, H.~Grigorian, A.~M.~\"Oztas and F.~Sandin,
  %``The Phase diagram of three-flavor quark matter under compact star constraints,''
  Phys.\ Rev.\ D {\bf 72}, 065020 (2005).
%  doi:10.1103/PhysRevD.72.065020
%  [hep-ph/0503194].
  %%CITATION = doi:10.1103/PhysRevD.72.065020;%%
  %174 citations counted in INSPIRE as of 14 Jul 2016
  
  %\cite{Prakash:1995uw}
\bibitem{Prakash:1995uw} 
  M.~Prakash, J.~R.~Cooke and J.~M.~Lattimer,
  %``Quark - hadron phase transition in protoneutron stars,''
  Phys.\ Rev.\ D {\bf 52}, 661 (1995).
%  doi:10.1103/PhysRevD.52.661
  %%CITATION = doi:10.1103/PhysRevD.52.661;%%
  %101 citations counted in INSPIRE as of 24 Jun 2016
  
  
  %%%%%%%%%%%%%%%%%%%%%%%%%%%%%%%%%%%%%%%%%%%%%
  
  %%%%%%%%%%%%%%%%%%%%%%%%%%%% hNJL
  
  %\cite{Benic:2014jia}
\bibitem{Benic:2014jia}
  S.~Benic, D.~Blaschke, D.~E.~Alvarez-Castillo, T.~Fischer and S.~Typel,
%{\it A new quark-hadron hybrid equation of state for astrophysics -
 %    I. High-mass twin compact stars},
  Astron.\ Astrophys.\  {\bf 577}, A40 (2015).
%  doi:10.1051/0004-6361/201425318
%  [arXiv:1411.2856 [astro-ph.HE]].
  %%CITATION = doi:10.1051/0004-6361/201425318;%%

%\cite{Benic:2014iaa}
\bibitem{Benic:2014iaa} 
  S.~Benic,
 % {\it Heavy hybrid stars from multi-quark interactions},
  Eur.\ Phys.\ J.\ A {\bf 50}, 111 (2014).
%  doi:10.1140/epja/i2014-14111-1
%  [arXiv:1401.5380 [nucl-th]].
  %%CITATION = doi:10.1140/epja/i2014-14111-1;%%
  
%%%%%%%%%%%%%%%%%%%%%%%%%%%%%%%%%%%%



  %%%%%%%%%%%%%%%%%%%%Non-local PNJL models
  
%\cite{Contrera:2012wj}
\bibitem{Contrera:2012wj} 
  G.~A.~Contrera, A.~G.~Grunfeld and D.~B.~Blaschke,
  %``Phase diagrams in nonlocal Polyakov-Nambu-Jona-Lasinio models constrained by lattice QCD results,''
  Phys.\ Part.\ Nucl.\ Lett.\  {\bf 11}, 342 (2014).
%  doi:10.1134/S1547477114040128
%  [arXiv:1207.4890 [hep-ph]].
  %%CITATION = doi:10.1134/S1547477114040128;%%
  
%\cite{Benic:2013eqa}
\bibitem{Benic:2013eqa} 
  S.~Benic, D.~Blaschke, G.~A.~Contrera and D.~Horvatic,
  %``Medium induced Lorentz symmetry breaking effects in nonlocal Polyakov–Nambu–Jona-Lasinio models,''
  Phys.\ Rev.\ D {\bf 89}, 016007 (2014).
  %doi:10.1103/PhysRevD.89.016007
  %[arXiv:1306.0588 [hep-ph]].
  %%CITATION = doi:10.1103/PhysRevD.89.016007;%%
  %17 citations counted in INSPIRE as of 06 Jul 2016
  
  
  \bibitem{Parappilly:2005ei}
  M.~B.~Parappilly {\it et al.}, 
%P.~O.~Bowman, U.~M.~Heller, D.~B.~Leinweber,
%A.~G.~Williams and J.~B.~Zhang,
  %``Scaling behavior of quark propagator in full QCD,''
  Phys.\ Rev.\ D {\bf 73}, 054504 (2006).
  %[arXiv:hep-lat/0511007].
  %%CITATION = HEP-LAT 0511007;%%

%\cite{Kamleh:2007ud}
\bibitem{Kamleh:2007ud} 
  W.~Kamleh {\it et al.}, 
%P.~O.~Bowman, D.~B.~Leinweber, A.~G.~Williams and J.~Zhang,
  %``Unquenching effects in the quark and gluon propagator,''
  Phys.\ Rev.\ D {\bf 76}, 094501 (2007).
%  [arXiv:0705.4129 [hep-lat]].
  %%CITATION = ARXIV:0705.4129;%%
  
  %\cite{Kaczmarek:2011zz}
\bibitem{Kaczmarek:2011zz} 
  O.~Kaczmarek {\it et al.}, 
  %F.~Karsch, E.~Laermann, C.~Miao, S.~Mukherjee, P.~Petreczky, 
  %C.~Schmidt and W.~Soeldner {\it et al.},
  %``Phase boundary for the chiral transition in (2+1) -flavor QCD at small 
  %values of the chemical potential,''
  Phys.\ Rev.\ D {\bf 83}, 014504 (2011).
%  [arXiv:1011.3130 [hep-lat]].
  %%CITATION = ARXIV:1011.3130;%%

    %\cite{Blaschke:2013rma}
\bibitem{Blaschke:2013rma} 
  D.~Blaschke, D.~E.~Alvarez Castillo, S.~Benic, G.~Contrera and R.~Lastowiecki,
  %``Nonlocal PNJL models and heavy hybrid stars,''
  PoS ConfinementX {\bf }, 249 (2012).
  %[arXiv:1302.6275 [hep-ph]].
  %%CITATION = ARXIV:1302.6275;%%
  %10 citations counted in INSPIRE as of 13 Jul 2016
  
      \bibitem{Blaschke:2013ana}
  D.~Blaschke, D.~E.~Alvarez-Castillo and S.~Benic,
  %{\it Mass-radius constraints for compact stars and a critical endpoint},
  PoS CPOD {2013} (2013), 063%; [{\tt arXiv:1310.3803 [nucl-th]}].
  %%CITATION = ARXIV:1310.3803;%%
  %8 citations counted in INSPIRE as of 22 Aug 2014
  
  %%%%%%%%%%%%%%%%%%%%%%%%%%%%
  %%%%%%%%%%%String-Flip Model %%%%%%%%%%%%%%%%%%%%%
  %\cite{Horowitz:1985tx}
\bibitem{Horowitz:1985tx} 
  C.~J.~Horowitz, E.~J.~Moniz and J.~W.~Negele,
  %``Hadron Structure In A Simple Model Of Quark / Nuclear Matter,''
  Phys.\ Rev.\ D {\bf 31}, 1689 (1985).
  %%CITATION = PHRVA,D31,1689;%%

%\cite{Ropke:1986qs}
\bibitem{Ropke:1986qs} 
  G.~R\"opke, D.~Blaschke and H.~Schulz,
  %``Pauli Quenching Effects in a Simple String Model of Quark / Nuclear Matter,''
  Phys.\ Rev.\ D {\bf 34}, 3499 (1986).
  %%CITATION = PHRVA,D34,3499;%%
  
  
  %\cite{Alvarez-Castillo:2015rwi}
\bibitem{Alvarez-Castillo:2015rwi} 
  D.~E.~Alvarez-Castillo, M.~A.~R.~Kaltenborn and D.~Blaschke,
  %``Excluded volume effects in the hybrid star EoS,''
  J.\ Phys.\ Conf.\ Ser.\  {\bf 668}, 012035 (2016).
  %doi:10.1088/1742-6596/668/1/012035
  %[arXiv:1511.05873 [nucl-th]].
  %%CITATION = doi:10.1088/1742-6596/668/1/012035;%%
  %2 citations counted in INSPIRE as of 06 Jul 2016
  
%\cite{Khvorostukin:2006aw}
\bibitem{Khvorostukin:2006aw} 
  A.~S.~Khvorostukin, V.~V.~Skokov, V.~D.~Toneev and K.~Redlich,
  %``Lattice QCD constraints on the nuclear equation of state,''
  Eur.\ Phys.\ J.\ C {\bf 48}, 531 (2006).
%  [nucl-th/0605069].
  %%CITATION = NUCL-TH/0605069;%%
   

%%%%%%%%%%%%%%%%%% Pasta Phases

   %\cite{Yasutake:2014oxa}
\bibitem{Yasutake:2014oxa} 
  N.~Yasutake, R.~Lastowiecki, S.~Benic, D.~Blaschke, T.~Maruyama and T.~Tatsumi. 
  %{\it Finite-size effects at the hadron-quark transition and heavy hybrid stars.}
  %``Finite-size effects at the hadron-quark transition and heavy hybrid stars,''
  Phys.\ Rev.\ C {\bf 89}, 065803 (2014).
  %; arXiv:1403.7492.
  %%CITATION = ARXIV:1403.7492;%%
  %3 citations counted in INSPIRE as of 14 Dec 2014
  
     %\cite{Alvarez-Castillo:2014dva}
\bibitem{Alvarez-Castillo:2014dva} 
  D.~E.~Alvarez-Castillo and D.~Blaschke,
 %{\it Mixed phase effects on high-mass twin stars},
  Phys.\ Part.\ Nucl.\  {\bf 46}, 846 (2015).
%  doi:10.1134/S1063779615050032
%  [arXiv:1412.8463 [astro-ph.HE]].
  %%CITATION = doi:10.1134/S1063779615050032;%%
 

  %%%%%%%%%%%%%%%%%%%%%%%%%%%%%%%%%%%%%%%%%%%%%%%%%%%%
  
  %\cite{Petran:2013qla}
\bibitem{Petran:2013qla} 
  M.~Petran and J.~Rafelski,
  %``Universal hadronization condition in heavy ion collisions at $\sqrt{s_\mathrm{NN}}= 62$ GeV and at $\sqrt{s_\mathrm{NN}}=2.76$ TeV,''
  Phys.\ Rev.\ C {\bf 88}, 021901 (2013).
 % doi:10.1103/PhysRevC.88.021901
 % [arXiv:1303.0913 [hep-ph]].
  %%CITATION = doi:10.1103/PhysRevC.88.021901;%%
  %23 citations counted in INSPIRE as of 30 Jun 2016
  
  %\cite{Rafelski:2014fqa}
\bibitem{Rafelski:2014fqa} 
  J.~Rafelski and M.~Petran,
  %``Strangeness in QGP: Hadronization Pressure,''
  Acta Phys.\ Polon.\ Supp.\  {\bf 7}, 35 (2014).
  %doi:10.5506/APhysPolBSupp.7.35
  %[arXiv:1403.4036 [nucl-th]].
  %%CITATION = doi:10.5506/APhysPolBSupp.7.35;%%
  %3 citations counted in INSPIRE as of 30 Jun 2016
  
%\cite{Suleimanov:2015fwa}
\bibitem{Suleimanov:2015fwa} 
  V.~F.~Suleimanov, J.~Poutanen, D.~Klochkov and K.~Werner,
  %``Measuring the basic parameters of neutron stars using model atmospheres,''
  Eur.\ Phys.\ J.\ A {\bf 52}, 20 (2016).
%  doi:10.1140/epja/i2016-16020-7
%  [arXiv:1510.06962 [astro-ph.HE]].
  %%CITATION = doi:10.1140/epja/i2016-16020-7;%%  
  
  %\cite{AlvarezCastillo:2012rf}
\bibitem{AlvarezCastillo:2012rf} 
  D.~E.~Alvarez-Castillo and S.~Kubis,
  %``Symmetry energy effects in the neutron star properties,''
  ASP Conf.\ Ser.\  {\bf 466}, 199 (2012).
  %[arXiv:1207.1078 [astro-ph.SR]].
  %%CITATION = ARXIV:1207.1078;%%
  %2 citations counted in INSPIRE as of 06 Jun 2016

   %\cite{Bazavov:2014pvz}
\bibitem{Bazavov:2014pvz} 
  A.~Bazavov {\it et al.}  [HotQCD Collaboration].% {\it The equation of state in (2+1)-flavor QCD.}
  %``Equation of state in ( 2+1 )-flavor QCD,''
  Phys.\ Rev.\ D {\bf 90}, 094503 (2014).
  %; arXiv:1407.6387.
  %%CITATION = ARXIV:1407.6387;%%
  %25 citations counted in INSPIRE as of 14 Dec 2014

  %\cite{Blaschke:2016lyx}
\bibitem{Blaschke:2016lyx} 
  D.~Blaschke, D.~E.~Alvarez-Castillo and T.~Kl\"ahn,
  %``Universal symmetry energy contribution to the neutron star equation of state,''
  arXiv:1604.08575 %[nucl-th].
  %%CITATION = ARXIV:1604.08575;%%
  %1 citations counted in INSPIRE as of 06 Jul 2016
  
  \bibitem{Bejger}
  M.~Bejger, {private communication}.
  
 \end{thebibliography}

\end{document}